\documentclass{iaus}
\usepackage{graphicx}

\newcommand{\apj}{ApJ}
\newcommand{\apjl}{ApJ}
\newcommand{\mnras}{MNRAS}
\newcommand{\aap}{A\&A}
\newcommand{\nat}{Nature}
\newcommand{\aj}{AJ}

\newcommand{\cm}{{\rm cm}}
\newcommand{\s}{{\rm s}}
\newcommand{\K}{{\rm K}}
\newcommand{\kms}{{\rm km}\,{\rm s}^{-1}}

\newcommand{\kpc}{{\rm kpc}}

\newcommand{\HI}{H$\,$\textsc{i}}
\newcommand{\HeII}{He$\,$\textsc{ii}}
\newcommand{\CIII}{C$\,$\textsc{iii}}
\newcommand{\CIV}{C$\,$\textsc{iv}}
\newcommand{\NV}{N$\,$\textsc{v}}
\newcommand{\OV}{O$\,$\textsc{v}}
\newcommand{\OVI}{O$\,$\textsc{vi}}
\newcommand{\SiIII}{Si$\,$\textsc{iii}}
\newcommand{\SiIV}{Si$\,$\textsc{iv}}

\newcommand{\lya}{Ly$\alpha$}

\title[Abundances in the high-$z$ IGM] 
{Abundances in the high-redshift Intergalactic Medium}

\author[Schaye \& Aguirre]   
{Joop Schaye$^1$ \and Anthony Aguirre$^2$%
}

\affiliation{$^1$Leiden Observatory, P.O. Box 9513, 2300 RA Leiden,
  The Netherlands \break email:
  schaye@strw.leidenuniv.nl\\[\affilskip] $^2$ UC Santa Cruz, 1156
  High Street, Santa Cruz, CA 95064, USA}

\pubyear{2005}
\volume{228}  
\pagerange{1--10}
\date{?? and in revised form ??}
\jname{From Lithium to Uranium: Elemental Tracers of Early Cosmic Evolution}
\editors{V. Hill, P. Fran\c{c}ois \& F. Primas, eds.}
\begin{document}

\maketitle

\begin{abstract}
The enrichment of the intergalactic medium (IGM) with heavy elements
provides us with a record of past star formation and with an opportunity to
study the interactions between galaxies and their environments. We
summarize current data analysis methods and observational constraints
on abundances in the diffuse, high-redshift ($z>2$) IGM. This
review is targeted at interested outsiders and attempts to answer the
following questions: Why should you care? 
What do we want to measure? How do we do it? What do we know? What are
the common misconceptions?

\end{abstract}

\firstsection 
\section{Introduction}

The enrichment of the tenuous gas in between galaxies
with elements heavier than helium, which are formed in
stars, provides us with an archaeological record of past star
formation and with a laboratory to study the interactions between
galaxies and their environments. 
The chemical composition of the intergalactic medium (IGM), which
contains most 
of the baryons in the universe and fills nearly all of space, is of
particular interest because it is linked to a number of issues central
to astrophysical cosmology.

The IGM comprises the baryonic reservoir from which galaxies form. The
rate at which the gas can dissipate energy as it undergoes
gravitational collapse is sensitive to the abundance of heavy
elements. The IGM metallicity therefore helps determine the rate of galaxy
formation as well as the masses of the stars that are formed. For
example, the collapse of metal-free gas may result in 
the formation of supermassive, so-called Population III stars, whereas
gas with a small amount of metals will form stars with an ordinary
mass function. 

It has long been realized that feedback from star formation, though
poorly understood, is key to the success of theories of galaxy
formation. 
Models of galaxies that do not include feedback processes,
form stars too efficiently and do not resemble observed galaxies.  
The energy and momentum that is deposited into the interstellar medium
by massive stars, is thought to drive supersonic outflows into
intergalactic space which carry heavy elements with them. We can
therefore study these galactic winds through the distribution of
metals in the IGM. 

The evolution of the IGM metallicity provides important constraints on
the cosmic star formation history, particularly at very high
redshift where the rate of star formation cannot yet be measured directly. 
Because
gravity and the expansion of the universe make it progressively more
difficult for outflows from galaxies to reach the low-density IGM,
regions away from galaxies are
promising places to search for the nucleosynthetic signature of
the first generations of stars. 
For example, relative abundances of different
heavy elements can be used to infer the initial mass function of the
stars that produced them. 

The ionization balance of heavy elements is
sensitive to the physical conditions in the gas and can be used to
constrain the gas density, temperature, and the spectrum of the
ionizing radiation to which it is exposed. Near galaxies that are known
to be driving winds, this kind of information can help us study the
physics of the outflows as well as of radiative feedback
processes. On larger scales, measurements of the spectral shape of the
background 
radiation can constrain the relative amounts of energy produced by
stars and by accretion onto black holes.

Given the wealth of information contained in IGM abundances, it is not
surprising that considerable efforts have been spent into mining this
information. In recent years progress has been particularly fast due to
several factors. First, the advent of high-resolution spectrographs on
ground-based 8m-class telescopes as well as on space telescopes, has
allowed us to obtain exquisite absorption spectra of a number of
quasars, which can fully resolve the \HI\ (though not
quite the metal) lines at a signal-to-noise of $\sim 10^2$. Since
each such spectrum allows identification of hundreds of absorption
features, much can be learned from relatively 
modest samples of quasars. Second, it has become clear that the
low-density IGM is governed by simple physics. On large scales 
the density fluctuations are only mildly
non-linear and the gas traces the dark matter, whereas on smaller
scales gas pressure needs to be taken 
into account. The relative simplicity of the physics governing the
low-density IGM allows us to make detailed theoretical predictions by
generating synthetic absorption spectra from large-scale, cosmological
simulations. Third, the availability of high-quality data and the need
to analyze vast numbers of realistic, synthetic spectra, has stimulated the
development of powerful, automated, statistical techniques that can be
used to study absorption that is weak compared to the noise (which for
many ions consists of contamination by other lines).

This review focuses on abundances in the diffuse IGM at redshift
$z>2$. For our purposes, Lyman limit systems (i.e., column 
densities $N_{\rm HI} > 10^{17}~\cm^{-2}$) 
are galaxies rather than intergalactic. These
systems are much more difficult to model than the diffuse IGM, but
they have the advantage that 
more elements can be detected. Abundances in these systems are reviewed
by Prochaska (this volume). 
This review provides neither a historical overview of the
subject, nor a complete list of references. 
Instead, I have opted to use the limited space available to
me to provide answers to a few basic questions that interested readers
with little specialized knowledge may have about abundances in the
IGM: Why should I care? What do we want to measure? How do we
do it? What do we know already (as of July 2005)? What are the common
misconceptions and  
which claims should I take with a grain of salt? The
first question was addressed in this section, the others will
come next.

\section{What do we want to measure?}

Depending on the enrichment mechanism, the abundances
of intergalactic metals could depend on various environmental factors,
such as the gas density, temperature, and the distance to particular
types of galaxies. The abundances are expected to be
stochastic, so we need to measure the probability distribution
function (pdf), not just the mean. Thus, we would like to
know, for each element, the pdf of its abundance as a function of
both time and environment. 

We would also like to know the fraction of the volume (i.e., the
volume filling factor) and of the mass that has an abundance higher than some
minimum value. Assuming we know the pdf of the density field, e.g.\
from theory, the filling factors can be obtained from the pdf of the
abundance as a function of density, which makes the density a
particularly interesting parameter.
 
\section{How do we do it?}
There are two fundamental problems that we need to overcome in order
to measure abundances in the diffuse IGM: the intergalactic gas has a
very low density, $n_{\rm H} \sim 10^{-6} - 10^{-3}~\cm^{-3}$ for $z
\lesssim 6$, and the abundances of heavy elements are typically very
low, $Z \lesssim 10^{-2}~Z_\odot$. The low density of heavy elements
has two important consequences. First, it is very difficult to detect
intergalactic heavy elements, particularly in the space-filling
low-density IGM. Second, the IGM is highly ionized by the
background radiation from galaxies and quasars, which means that
we need to know the ionization balance in order to convert ion
abundances into heavy element abundances.

\subsection{Strategy}
To (partly) overcome the detection problem, the community has adopted
the strategy outlined below. 

\begin{enumerate} 
\item \emph{Use absorption rather than
emission}. The optical depth is proportional to the density, whereas
the emissivity scales as the density squared. Hence, absorption
studies are much less biased to high-density
regions. The great disadvantage of absorption studies is
the need for a bright background source. The background sources of
choice are the most luminous quasars, because these are the brightest
sources that can be observed over cosmological distances. Because
quasars are effectively point sources, this does have the unfortunate
consequence that the information is one-dimensional.

\item \emph{Use lots of time on the largest
telescopes}. Using 8m class telescopes, it is currently feasible to
obtain spectra with a resolution 
$\Delta v \sim 10~\kms$ ($R\sim 30,000$) and a signal-to-noise ratio
${\rm S}/{\rm N} \sim 10^2$. Comparison of this resolution with the thermal
line width,
\begin{equation}
b \equiv \left ( {2kT \over m}\right )^{1/2} \approx 12.8 \left
  ({m_{\rm H} \over m} \right )^{1/2}\left ({T 
  \over 10^4\,\K} \right )^{1/2}~\kms,
\label{eq:b}
\end{equation}
where $T\sim 10^4~\K$ is the natural temperature for a photo-ionized 
hydrogen/helium plasma, shows that \HI\ lines can be fully resolved,
but that the metal lines are typically instrumentally broadened.

\item \emph{Observe in the rest-frame ultraviolet}, which is where
many of the  
strongest transitions lie of the ions of interest. If we want to use
ground-based telescopes, this limits us to the redshift range $z\sim
1.5 - 6$ with $z\sim 2.5 - 3.5$ being optimal. Coincidentally, for
this redshift range the information content per quasar spectrum is close
to maximum because the \HI\ \lya\ forest is already thick enough to produce
many absorption lines per unit redshift, yet still thin enough that most of
these lines are not saturated (note that the thinning of the forest is
mostly driven by the universal expansion). Fortunately, this redshift
range is also 
intrinsically very interesting because it corresponds to the epoch at
which quasar activity and the star formation rate were (close to)
maximum. Given these lucky coincidences, it is perhaps not surprising
that most of the work on IGM abundances has focused on $z\sim 2 -
4$, which is why this review focuses on this epoch. This is not to say
that lower and higher redshift work is uninteresting. Observations at
$z<1.5$ (which need to be done from space) are for example important
because they cover most of cosmic history and observations at $z>5$
are important because they can tell us about reionization and the
first generations of galaxies.

\item \emph{Focus on the transitions that can be detected most
easily} because they are strong, occur in ions that are abundant, are
part of a multiplet (which eases identification),
and/or fall in regions of the spectrum that are relatively free from
contamination. Since most of the absorption lines are due to \HI\
along the line of sight to the quasar, the
level of contamination worsens progressively when we move from
rest-frame wavelength $\lambda > \lambda_{{\rm Ly}\alpha}$ to
$\lambda_{{\rm Ly}\beta} <  \lambda < \lambda_{{\rm Ly}\alpha}$ to 
$\lambda_{{\rm LL}} <  \lambda < \lambda_{{\rm Ly}\beta}$ to 
$\lambda < \lambda_{{\rm LL}}$, where $(\lambda_{\rm LL},\lambda_{{\rm
Ly}\beta}, \lambda_{{\rm Ly}\alpha}) \approx (912,1026,1216)$~\AA. The
metal transitions that have so far been studied (in more than a few
systems) are, in order of ease of detection, \CIV\ (1548, 1551), \SiIV\
(1394, 1403), \OVI\ (1032, 1038), \CIII\ (977), and \SiIII\
(1207). 

\item \emph{Aim to detect metal absorption statistically rather than by
eye}. Traditionally, absorption line systems have been identified by eye,
and then deblended into Voigt profile components characterized by their
redshift $z$, line width $b$, and column density $N$. This approach
is well-suited for strong absorbers that are relatively free from
contamination, but too insensitive and time-consuming for
the analysis of large data sets and for studying
absorption that is weak compared to the noise/contamination. 
The statistical technique that has been used most widely, namely the
so-called pixel optical depth (POD) technique, will be described in
\S\ref{sec:pod}. 

\item \emph{Use synthetic absorption spectra drawn from large-scale
hydrodynamical simulations to test methods and to help guide the
interpretation}. This approach was first applied to the
\HI\ \lya\ forest and has been instrumental for the acceptance
and development of the current physical picture of the forest. In
recent years, studies of intergalactic metals have used cosmological
simulations to test methods, to measure abundances by comparing
observed spectra to simulated ones,
and to make ab-initio predictions. Given
the complexity of 
searching for a weak signal, which can be due to gas with varying
physical properties, in noisy data, the use of simulations to make
measurements is often essential. However, it is important to keep in mind
that this approach has the disadvantage that the measurements are
subject to the numerical and physical validity of the
simulations.
\end{enumerate}

\subsection{The pixel optical depth technique}
\label{sec:pod}

The pixel optical depth (POD) technique was pioneered by
Cowie \& Songaila (1998) and developed extensively by
Aguirre, Schaye, \& Theuns (2002). It is fast, 
sensitive, and 
automated, making it particularly well-suited for the analysis of (both
weak and strong) absorption in large data sets and simulations.
Its main drawback is that the interpretation of
the results is not straightforward when the absorption is weak, an
issue that I will come back to in \S\ref{sec:percentiles}.

The key steps in the POD search are:
\begin{enumerate} 
\item Compute arrays of apparent pixel optical depths,
\begin{equation} 
\tau_{\rm app}(z) = -\ln{F(z) \over F_{\rm cont}(z)},
\end{equation}
where $F$ and $F_{\rm cont}$ are the transmitted flux in the pixel and
the local continuum, respectively. We refer to these PODs as apparent
because they may be 
affected by contamination, noise, continuum fitting errors,
etc. Sophisticated methods have been developed to recover better
estimates of the true optical depths. These methods take advantage of the
fact that many ions 
generate multiplets (e.g., \HI) or doublets (e.g., \CIV, \NV, \OVI,
and \SiIV), with known optical depth ratios between the
transitions. Algorithms have been developed to reject strong
contamination (Aguirre et al.\ 2002), correct
for saturation (relevant for \HI; Cowie \& Songaila 1998; Aguirre et
al. 2002),  
correct for contamination by higher order \HI\ Lyman lines (relevant
for \CIII\ and \OVI; Aguirre et al.\ 2002), and to correct for
self-contamination (relevant for \CIV\ and \SiIV; Aguirre et al.\ 2002).
\item Choose a ``base'' and a ``target'' transition and bin the pixel pairs
  according to the recovered POD of the former (e.g., \CIV\
  as a function of \HI). The method
  is most sensitive if the two transitions occur in similar gas phases
  and if the base transition is more easily detectable than the target
  transition. The
  combinations that have so far proved to be most useful are
  \CIV(\HI) (Cowie et al.\ 1998), \OVI(\HI) (Schaye et al.\ 2000),
  \SiIV(\CIV) (Aguirre et al.\ 2004), \CIII(\CIV) (Schaye et al.\
  2003), and \SiIII(\SiIV) (Aguirre et al. 2004).  
\item Compute the median (or any other percentile) of the target transition
  POD as a function of the binned base transition POD. Errors can for
  example be computed by cutting the redshift range in chunks (that
  are larger than the scale on which PODs are correlated) and
  bootstrap resampling the chunks. A correlation between the 
  median target PODs and the base optical depth reflects a
  detection\footnote{Note 
  that the absence of a correlation does not necessarily imply that
  the contribution of the target transition to the
  recovered optical depths is insignificant.}. The use of
  non-parametric statistics (such as the median) increases the
  robustness of the method, which is 
  important given the imperfect recovery of the PODs.
  The method can be generalized to measure the full distribution of
  target optical depths by simultaneously measuring multiple
  percentiles (Schaye et al.\ 2003).

\item Determine the noise level (which includes contamination) and,
  optionally, 
  subtract its contribution to the optical depth. For example, if
  the median $\tau_{\rm CIV} \rightarrow C$ as $\tau_{\rm HI} \rightarrow 0$,
  then this indicates that $C$ is the median noise level. We can
  attempt to correct for the noise contribution by subtracting $C$
  from the median target PODs. Note, however, that the
  validity of this correction depends on the distribution of the true
  optical depths (it is only strictly valid if the
  distribution of true PODs is much narrower than that of the noise, see also
  \S\ref{sec:percentiles}). 

\end{enumerate}

\subsection{Ionization corrections}
We want to know elemental abundances, but we can only measure
ion abundances. This
is both a curse and a blessing. It is a curse because we need to
correct for ionization if we are interested in the abundances of heavy
elements. Uncertainties in the ionization balance are currently the
limiting factor in the study of intergalactic abundances. However,
the fact that the observables depend on the ionization balance is also
a blessing because it 
means that we can constrain the physical conditions in the gas and
that we can measure abundances for different gas phases.

The ionization balance depends in general on the following parameters:
\begin{enumerate}
\item The \emph{radiation field}. It is helpful to split this free
  function into an overall normalization and a spectral shape. If, as
  is usually done, we assume that the radiation field is uniform, then
  the normalization is measurable in the form of the \HI\ ionization
  rate (which determines the mean \HI\ \lya\
  absorption). The spectral hardness, which depends mainly on the
  relative contributions of stars and quasars, as well as on the
  ionization balance of helium, is currently the main source of
  uncertainty in studies of intergalactic abundances. We either have
  to choose a model UV background (based on observations of galaxies and
  quasars), in which case we can measure relative abundances, or we
  have to assume that we know the relative abundances in which case we can
  constrain the spectral shape. It is, however, possible to get
  interesting constraints on both if we allow for some weak
  priors. For example, Aguirre et al.\ (2004) find that the prior
  ${\rm Si/C} < 10 {\rm (Si/C)}_\odot$ rules out a UV 
  background dominated by quasars, while even extremely soft
  backgrounds result in an overabundance of Si relative to
  C. Naturally, by observing more ions, we can draw
  interesting conclusions using fewer and/or weaker assumptions.

\item The \emph{gas density}. We can measure the gas density from
  the \HI\ optical depth since $n_{\rm HI} \propto \rho^2 \Gamma_{\rm
  HI}^{-1}$ in 
  photo-ionization equilibrium and the \HI\ ionization
  rate $\Gamma_{\rm HI}$ can be measured. If we
  use $\tau_{\rm HI}$ as the base optical depth, we can convert a
  plot of $\tau_i(\tau_{\rm HI})$ into a plot of the corresponding
  elemental abundance as a function of density, $Z_i(\rho)$.
  We can perform consistency checks because ratios like \CIII/\CIV\
  and \SiIII/\SiIV\ are sensitive to the gas density in
  photo-ionization equilibrium (Schaye et al.\ 2003; Aguirre et al.\
  2004).  

\item The \emph{temperature}. If the gas is very hot and/or dense,
  then collisional ionization may dominate over photo-ionization, in
  which case the ionization balance will depend only on the
  temperature. If photo-ionization dominates, then the ion fractions
  are still weakly dependent on the temperature because the recombination
  rates are. However, as long as $T\sim 10^4~\K$, as is usually assumed,
  the temperature is not the main source of uncertainty in the
  analysis. 

  There are several ways in which we can constrain the
  temperature from above. Line widths are most widely used, but
  the spectral resolution of current observations is generally
  insufficient to rule out collisional ionization of heavier elements
  such as Si.
  Ratios like \CIII/\CIV\ and \SiIII/\SiIV\ can be used to
  rule out temperatures sufficient for collisional ionization to
  dominate (Schaye et al.\ 2003; Aguirre et al.\ 2004). 

\item The ionization \emph{history}. It is usually assumed that the
  gas is in ionization equilibrium. Without this assumption it is
  impossible to correct for ionization because we would need to know
  the history of the gas to compute the ionization balance.
  Ionization equilibrium should be an excellent
  approximation for hydrogen, because the photo-ionization timescale
  is only of order ten thousand years. However, for the ions
  with ionization energies that exceed that of hydrogen the relevant
  time scales can be much greater. Moreover, if heavy elements are
  carried by supersonic flows, then much of the enriched gas may be
  shocked to high temperatures and subsequent rapid cooling could then
  give rise to deviations from ionization equilibrium. 
\end{enumerate}

Note that it is likely that all of the assumptions on which the
ionization corrections are based, break down near galaxies with high
star formation rates or active nuclei. For example, radiation from
local sources is likely to be important for many of the stronger metal
line systems (Schaye 2004).

\section{What do we know?}

Before summarizing recent results on abundances at $z> 2$, it is
useful to look at three equations that can help us interpret the
observational results. Assuming local hydrostatic equilibrium, which
is likely to be a good approximation for overdense gas, and
photo-ionization equilibrium, we can 
relate the \HI\ column density to the gas density (Schaye 2001):
\begin{equation}
N_{\rm HI} \sim 10^{13.5}~\cm^{-2} \left ({\rho \over \left
  <\rho \right >}\right )^{3/2} \left ({1+z \over 4}\right )^{9/2}
  \left ({\Gamma_{\rm HI} \over
  10^{-12}~\s^{-1}} \right )^{-1} \left ({T \over 10^4~\K}\right )^{-0.26} 
\end{equation}
For a thermal line profile, $\tau(v) = N {c\sigma \over \sqrt{\pi}b}
e^{-v^2/b^2}$, where $c$ and $\sigma$ denote the speed of light and
the cross-section for absorption respectively, the optical depth at
the line center is given by  
\begin{equation}
\tau_{\rm c} = \left ({f\lambda_0 \over
  f_{{\rm Ly}\alpha}\lambda_{0,{\rm Ly}\alpha}}\right ) \left ({N \over
  10^{13.5}~\cm^{-1}}\right ) \left ({b \over 24~\kms}\right )^{-1}, 
\end{equation}
where $f$ and $\lambda_0$ are the oscillator strength and rest
wavelength, respectively. For gas with densities below the
cosmic mean, pressure forces play no role because the Jeans scale
exceeds the sound horizon. Therefore, the absorption due to gas of
very low densities is better described by the fluctuating
Gunn-Peterson approximation,
\begin{equation}
\tau_{\rm HI,GP} \sim \left ({\rho \over \left
  <\rho \right >}\right )^2 \left ({1+z \over 4}\right )^{9/2}
  \left ({\Gamma_{\rm HI} \over
  10^{-12}~\s^{-1}} \right )^{-1} \left ({T \over 10^4~\K}\right
  )^{-0.76}.  
\end{equation}

Soon after the advent of the HIRES echelle spectrograph at the Keck
telescope, it became clear that for about half of all absorbers with
$N_{\rm HI} > 10^{14.5}~\cm^{-2}$ the associated \CIV\ is
detectable with $N_{\rm CIV}/N_{\rm HI} \sim 10^{-2.5}$ (Cowie et al.\
1995). The highest quality spectra available
(S/N$\sim 300$ for \CIV), show no evidence for a turnover in the
column density distribution down to $N_{\rm
  CIV} \sim 10^{11.7}~\cm^{-2}$ (Ellison et al.\ 2000).
Using pixel optical depth techniques (see \S\ref{sec:pod}),
it is possible to obtain statistical detections of \CIV\ down to
$\tau_{\rm HI} \sim 1$ (e.g., Cowie \& Songaila 1998;
Ellison et al.\ 2000; Schaye et al.\ 2003; Aracil et al.\ 2004) and
down to somewhat lower \HI\ optical depths for \OVI\ (e.g., Schaye et al.\
2000; Aracil et al.\ 2004, but see Pieri \& Haehnelt 2004). However, it is
unclear what fraction of these low-$\tau_{\rm HI}$ pixels has
associated metal absorption (see \S\ref{sec:percentiles}).

\begin{figure} 
\resizebox{\textwidth}{!}{\includegraphics{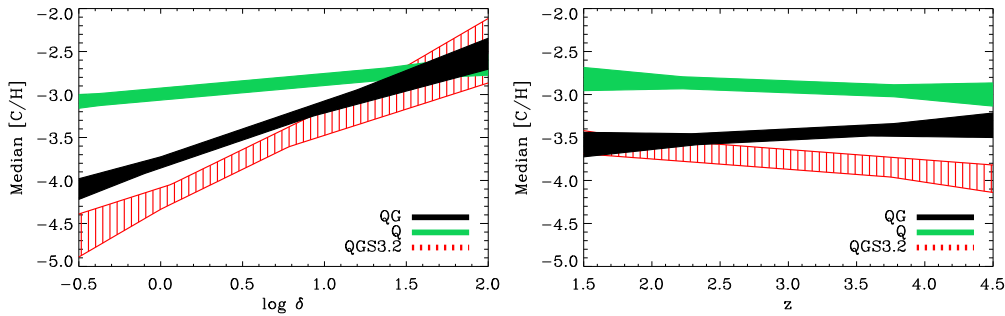}}
\caption{Shaded regions enclose the $1\sigma$ confidence contours for
  fits to the carbon abundance as a function of overdensity $\delta
  \equiv \rho/\left < \rho\right > -1$ and
  redshift for $z=3$ (\emph{left}) and $\log\delta = 0.5$
  (\emph{right}). Results are shown for the fiducial UV background
  model (QG) for galaxies and quasars, as well as for a pure quasar
  background (Q) and for a model in which the flux was reduced by a
  factor 10 above 4~Ryd to mimic incomplete \HeII\ reionization
  (QGS3.2). Figure taken from Schaye et al.\ (2003).
\label{fig:results}
} 
\end{figure}

Both simple
photo-ionization models and comparisons with 
hydrodynamical simulations assuming a uniform metallicity, have shown that
the observations imply a low IGM metallicity, $Z \lesssim 10^{-2}
Z_\odot$ (e.g., Cowie et al.\ 1995; Haehnelt 
et al.\ 1996; Songaila \& Cowie 1996; Hellsten et al.\ 1997; Rauch et
al.\ 1997; Dav\'e et al.\ 1998). Recent studies have dropped the
assumption of a uniform metallicity and used \HI\ dependent
ionization corrections to recover the pdf of the metallicity
as a function of density from large samples of quasar
spectra. Schaye et al.\ (2003) used the POD technique
and estimated densities using calibrated hydrodynamical
simulations. They found that the abundance of carbon is a strongly
increasing function of density, but not of time (see
Fig.~\ref{fig:results}). The size of the 
gradient is sensitive to the hardness of the UV background, but the
conclusion that evolution is weak between $z=4$ and 2 is
not. Integrating over the full distribution, they found that the
contribution of photo-ionized gas with density $-0.5 < \log \rho/\left <
\rho\right > < 2.0$ to the cosmic carbon abundance is $[{\rm C}/{\rm
    H}] = -2.8 \pm 0.1$ for their fiducial UV background from galaxies
and quasars, and about 0.5
dex higher for a pure quasar background. Independent of the model for 
the UV background, they found that the distribution is highly
inhomogeneous and can be fit by a lognormal distribution of width
$\sim 0.75$~dex. These results were confirmed by
Simcoe et al.\ (2004), who used Voigt 
profile decompositions rather than PODs and who estimated
the densities using the analytic model of Schaye (2001) instead of
hydrodynamical simulations. 

Metal lines are clustered on scales up to $\sim 10^3~\kms$ (e.g.,
Pichon et al.\ 2003; Boksenberg et al.\ 2003) and their
autocorrelation exceeds that expected for a metal distribution that
depends only on the density (Scannapieco et al.\ 2005). Adelberger et
al.\ (2003, 2005) found that the cross-correlation between \CIV\ lines
and Lyman-break galaxies increases with $N_{\rm CIV}$ and becomes
comparable to the galaxy autocorrelation for $N_{\rm CIV} \gtrsim
10^{12.5}~\cm^{-2}$, which suggests that the strongest \CIV\ lines arise
in gas that is directly associated with the galaxies. 
Motivated by these findings, Pieri, Schaye, \& Aguirre (2005) showed
that the abundances of C and O depend on both the gas density
and the proximity to highly-enriched regions (and thus
galaxies). Pieri et al.\
also found that the parameter ``galaxy proximity'' can
account for part of the scatter in the abundance of carbon at fixed
$\tau_{\rm HI}$.  

Songaila (2001) found no evidence for evolution in the column density
distributions of \CIV\ and \SiIV\ from $z\approx 5$ to 2, although the
constraints are poor for $z>4$. The
integral of the distribution is insensitive to weak lines and was
found to be roughly constant over the same redshift range (Songaila
2001; Pettini et al.\ 2003; but see Songaila 2005). Although these
observations cannot tell us about the evolution of the elemental
abundances (see \S\ref{sec:evolution}), they do provide a lower
limit to the metallicity of the universe at $z\approx 5$ of $[{\rm
    C}/{\rm H}] \approx -4$. 

Aguirre et al.\ (2004) compared POD statistics of observed spectra
with those of
spectra drawn from hydro simulations that use the carbon
distribution measured by Schaye et al.\ (2003), but a variable Si/C
ratio. They found that Si is overabundant relative to C and that a
single Si/C ratio can reproduce the full distribution of \SiIV/\CIV. 
The exact overabundance of Si is sensitive to the hardness of the UV
background. It becomes unphysically high for a quasar background, and
remains $\gtrsim 0.5$~dex for a background dominated by galaxies. For
a soft background, and according to an analysis of \OV\ even for
a quasar background (Telfer et al.\ 2002), O is also inferred to be
overabundant relative to C (Bergeron et al.\ 2002; Simcoe et al.\
2004). 

Using the POD technique it is possible to detect \CIII\ and \SiIII\
(Songaila 1998). Schaye et al.\ (2003) and Aguirre et al.\ (2004)
demonstrated that the observed ratios \CIII/\CIV\ and \SiIII/\SiIV\
rule out temperatures high enough for collisional ionization to
dominate, but agree with the predictions for warm, photo-ionized
gas. These ratios are also sensitive to the density of photo-ionized
gas and the agreement between predictions and observations confirms
that the \HI\ absorption arises in the same gas phase as the
associated metal absorption. For \OVI, on the other hand, the
situation is more complicated. At $z\lesssim 2.5$, \OVI\ is detectable
in a large fraction of absorbers with $N_{\rm HI} \gtrsim
10^{14}~\cm^{-2}$ and for many of these the observed line widths rule
out collisional ionization (Carswell et al.\ 2002;
Bergeron et al.\ 2002). However, the much rarer \OVI\ systems with
$N_{\rm HI} > 10^{15}~\cm^{-2}$ often show evidence for a complex,
multiphase structure (e.g., Carswell et al.\ 2002, Simcoe et al.\
2002), and the \OVI\ line widths are suggestive of collisional
ionization ($T \gtrsim 10^5~\K$, Simcoe et al.\ 2002). Interestingly, a
fraction of the \OVI\ associated with low $N_{\rm HI}$ 
absorbers appear to be metal-rich and too compact to be
self-gravitating (Carswell et al.\ 2002; Bergeron et al.\ 2002). 

\section{What are the common misconceptions?}

In this section I will discuss a number of common interpretations and
claims that, in my opinion, deserve more scepticism than is generally
appreciated.

\subsection{Gas density}
The density at a given
point in space depends on the smoothing scale that is used. For
example, the density at the point that coincides with the
location of this page, will differ enormously depending on whether we
smooth over a scale of 1~cm, 1~AU, or 1~Mpc. Therefore,
density and volume filling factors are meaningless unless we specify a
smoothing scale. Moreover, unless the heavy and light elements are
fully mixed, the density of metals and the density of the gas will not
have the same dependence on smoothing scale. 
Unfortunately, smoothing scales are almost never specified, thereby
severely limiting the potential usefulness of much of the theoretical
and observational literature. 

Studies that estimate the density
from the column density of \emph{fully-resolved} \HI\ \lya\
lines, effectively smooth their densities on scales corresponding to
the sizes of the absorbers, which, for
overdense absorbers, should typically be close to the local Jeans
length and will therefore depend on the density
itself (Schaye 2001):
\begin{equation}
L \sim 1.0 \times 10^2~\kpc \left ({N_{\rm HI} \over 10^{14}~\cm^{-2}}
\right )^{-1/3} \left ({T \over 10^4~\K}\right )^{0.41} \left
({\Gamma_{\rm HI} \over 10^{-12}~\s^{-1}}\right )^{-1/3}. 
\end{equation}
Studies that use the \HI\ \lya\ optical
depth to estimate the density will effectively smooth the density
field twice: once as for column densities and then again on the
thermal broadening scale ($\sim 2b/H(z)$). Using equation (\ref{eq:b})
and the high-redshift approximation $H\propto (1+z)^{3/2}$ for the
Hubble parameter, we obtain
\begin{equation} 
L_{\rm th} \sim 58~h^{-1}\,\kpc~\left ({m_{\rm H} \over m} \right
  )^{1/2}\left ({T \over 10^4\,\K} \right )^{1/2} \left ({1+z \over
  4}\right )^{-3/2}. 
\end{equation}
Note that in this case the smoothing length depends on the mass of the
ion being ionized, which complicates the interpretation. One way to
solve this problem is to smooth the spectrum on the thermal scale of
the lightest element (e.g., Aguirre et al.\ 2004). Naturally, if the
\HI\ lines are not resolved, then the smoothing scale will be set by
the spectral resolution. 

\subsection{Evolution}
\label{sec:evolution}
Observations suggest that the global densities of \CIV\ (i.e.,
$\Omega_{\rm CIV}$) and \SiIV\ have not 
evolved significantly from $z=5$ to $z=2$ (Songaila 2001: Pettini et
al.\ 2003, but see Songaila 2005). These observations are widely
quoted to support the idea 
that the IGM was enriched by small galaxies at very high
redshift. However, there are two reasons why the evolution of
$\Omega_{\rm CIV}$ (and similarly $\Omega_{\rm SiIV}$) does not tell
us much about the evolution of the 
IGM metallicity. First, the integral over the observed \CIV\ column
density distribution diverges at high $N_{\rm CIV}$ if we extrapolate
to high column densities. Thus, very large quasar samples are
needed to constrain $\Omega_{\rm CIV}$. Moreover, even if such samples
were available, they would probe the evolution of the number density of
the strongest \CIV\ systems, which are usually associated with Lyman
limit systems rather than the diffuse IGM. Second,
ionization corrections can be as large as a factor 10 (Schaye et al.\
2003). If, as expected for the diffuse IGM, these corrections depend on
redshift, then the evolution of $\Omega_{\rm CIV}$ will tell us very
little about the evolution of $\Omega_{\rm C}$. 

Schaye et al.\ (2003) did correct for ionization and did not
find any evidence for evolution in the distribution of carbon in the
diffuse IGM from
$z\approx 4$ to 2 for any of their UV background models. However, at
$2\sigma$ confidence their results 
still allow a factor of 2 increase in the abundance of
carbon. Considering that the age of the universe roughly doubles 
from $z=4$ to 2, it is clear that better measurements are needed
before we can draw strong conclusions about the epoch of IGM enrichment.

\subsection{Percentiles}
\label{sec:percentiles}
Suppose that the median \CIV\ optical depth or column density
corresponding to some \HI\ bin
significantly exceeds the median noise/contamination
level, i.e., carbon is
detected. What does this imply for the fraction of the absorbers (with the
corresponding \HI\ strength) that are enriched? Nearly all studies in
the literature have assumed either that all are enriched to this median
level, or that 50\% of the absorbers are enriched to a level greater
than the median. The truth is, however, that we simply do not know. It
is possible that the common interpretations are correct, but it is also
possible that only a small fraction of the absorbers contain
carbon. 

Take, for example, a sample of
100 absorbers, all but one of which are metal-free. The median metal
abundance will in that case be consistent with zero. Now take a sample
of $10^6$ absorbers of which 99\% are again metal-free. The median can
in that case be measured precisely and may
significantly exceed the median noise level,
but it would clearly be wrong to conclude that at least 50\% of the
absorbers are enriched. On the other hand, if the observed median
\CIV\ strength would exceed
the 99.9th percentile of the noise pdf, then the conclusion that at
least 50\% of the absorbers contain carbon would be justified. These
examples show that it is possible to recover information on the
filling factor of metals, but that this requires a very careful
analysis if the observed metal abundances are close to the detection
limit. Up till now, such an analysis has not been carried
out and consequently we have no strong constraints on the filling
factor of enriched, low-density gas (which dominates the overall
filling factor).

\subsection{Incompleteness corrections}
Observed distributions of column
densities are often ``corrected'' for incompleteness. For example,
Monte Carlo simulations 
have been used to estimate the amount of metal absorption in the form
of lines (e.g., Ellison et al.\ 2000) and pixels (e.g., Songaila
2005) that has been missed due to noise and/or line blending. This
procedure is arbitrary because one has to decide down to which column
density incompleteness corrections are applied. It is also model-dependent
because one has to assume an input distribution for the Monte Carlo
simulations. Thus, it is important to check whether observational
results have been ``corrected'' for incompleteness or whether they
represent true detections.

\subsection{Metal inventory} 
It is necessary to know the ionization
balance in order to convert observed ion abundances into 
elemental abundances. However, metals residing in gas which is very
hot, $T >> 10^5~\K$, will be so highly ionized that they are
essentially invisible in rest-frame ultraviolet absorption
spectra. It is therefore important to note
that current observational constraints apply only to the warm, $T\sim
10^4 - 10^5~\K$, IGM. We cannot rule out the possibility that
significant amounts of intergalactic metals are hidden in hot,
collisionally ionized gas.

\section{Conclusions}
\label{sec:concl}
The low densities and metallicities that are characteristic of the
diffuse IGM make it 
very difficult to detect heavy elements. Once detected, the
interpretation is complicated by the large ionization corrections that
are needed to convert observed ion abundances into
elemental abundances. Despite these challenges, tremendous progress has been 
made, driven by the quality of
the quasar spectra delivered by echelle 
spectrographs on 8m-class telescopes, the availability of
increasingly realistic hydrodynamical simulations, and the development
of sophisticated statistical techniques.

Analyses of absorption by \HI,
\CIII, \CIV, \OVI, \SiIII, and \SiIV\ have revealed that the
distribution of metals is highly inhomogeneous, that the metallicity
of the IGM is a strong function of the environmental parameters
density and proximity to galaxies, that there is not much room for
evolution from $z\sim 4$ to 2, and that a substantial reservoir of
metals was already in place by $z = 5$. We also know that Si, and probably
also O, is highly overabundant relative to carbon.
While most of the enriched gas clouds
are found to be photo-ionized and metal-poor, there is some
evidence for hot, collisionally ionized oxygen, particularly at high
densities, and there exist some metal-rich clouds which are too
compact to be self-gravitating. 

Although we have already learned a lot, it is clear that we have only just
begun to extract the wealth of information contained even in existing
observations. As methods, models, and instruments continue to improve, 
so will our understanding of the chemical composition of the universe.

\begin{acknowledgments}
I am grateful to the organizers for inviting me to a great conference. 
This work was supported by a Marie Curie Excellence
Grant from the European Union (MEXT-CT-2004-014112). 
\end{acknowledgments}

\end{document}